\documentstyle[epsfig]{aipproc}

\begin{document}

\title{Correlated Magnetoexcitons in Semiconductor 
       Quantum Dots at Finite Temperature}

\author{
           D.J. Dean$^\dagger$, 
           M.R. Strayer$^\dagger$, and 
           J.C. Wells$^*$
\\
$^\dagger$Physics Division and 
$^*$Computer Science and Mathematics Division, 
\\
Oak Ridge National Laboratory, Oak Ridge, TN 37831
}

 \maketitle

\vspace{0.2in}
\noindent
{\large\bf ABSTRACT}
\vspace{0.2in}

We describe computational methods for the theoretical 
study of explicit correlations beyond the mean field 
in excitons confined in semiconductor quantum dots in 
terms of the Auxiliary-Field Monte Carlo (AFMC) 
method\cite{kdl97}.
Using AFMC, the many-body problem is formulated as a 
Feynman path integral at finite temperatures and 
evaluated to numerical precision.  This approach is 
ideally suited for implementation on high-performance 
parallel computers.  Our strategy is to generate a 
set of mean-field states via the Hartree-Fock method 
for use as a basis for the AFMC calculations.  
We present preliminary results.

\vspace{0.2in}
\noindent
{\large\bf INTRODUCTION}
\vspace{0.2in}

Revolutionary nanofabrication techniques, such as atomic-layer
epitaxy, advanced lithography, and electrochemical and molecular
self-assembly methods, have opened the way to fabricating
quantum dots, arrays of quantum dots, and other structures
that may provide the basic building blocks for future nano-scale
electro-optic devices \cite{Jacak}.
Recent experimental studies of optical emission from
semiconductor quantum dots have demonstrated the importance
of many-body effects, such as correlation and exchange,
in exciton dynamics \cite{Landin}.
Intensive basic research is ongoing to accurately simulate the 
complex properties of current and future nanostructure devices 
using high-performance computers (e.g., see \cite{Iafrate}).
While significant progress has been made in the simulation of the
relevant independent-particle phenomena,
much important work remains in understanding and simulating
the collective and correlated many-body phenomena observed in
nanoscale devices at finite temperature \cite{Hess}.
Theoretical research has focused largely on mean-field or one-body 
aspects of the phenomena at zero temperature.
Many-body effects, such as exciton-exciton
interactions, are still lacking a complete theoretical
understanding and a numerical implementation.

The focus of this paper is the theoretical description and
exact, fully correlated numerical simulation of many-electron
correlations in two-dimensional semiconductor quantum dots 
at finite temperature and exposed to strong magnetic fields.
We will approach this many-body problem via the Auxiliary-Field
Monte-Carlo (AFMC) method \cite{SK85},
which has been known since the 1950s \cite{hstran},
but without wide application to fermion systems because of
difficulties in numerical implementations known as the
``fermion-sign'' problem.  Nuclear physicists have pioneered 
the modern development of AFMC for the shell-model problem \cite{kdl97},
obtaining practical solutions to the associated numerical issues.
With this important work as inspiration, AFMC has been applied
in condensed-matter systems \cite{Linden}
and atomic and molecular physics \cite{Neuhauser}.
In those cases for which practical solutions to the sign problem
have been found, the results have been quite revolutionary for
the theoretical description of the phenomena.
No one, to our knowledge, has tried to apply the AFMC method to
quantum nanostructures.
This new effort will provide a numerically exact treatment of the
quantum many-body problem in nanostructures at finite temperature.

In describing the ground-state and low-lying (intraband) excitations of
the N-electron semiconductor nanostructures, it is often sufficient to
restrict consideration to the conduction band using the effective-mass
approximation\cite{Jacak}.
Denoting the operators of creation and annihilation of an electron
in a single-particle state
$| \alpha \rangle $ by $ c_\alpha^{\dag} $ and $ c_\alpha $,
respectively, the Hamiltonian of the N-electron system can be
written in the occupation-number representation in the form
\begin{equation}
\hat{H}_{ee}=
\sum_\alpha E^{e}_\alpha \hat{c}_\alpha^{\dag} \hat{c}_\alpha
          + \frac{1}{2} \sum_{\alpha \beta \gamma \delta}
            \langle \alpha \beta | V_{ee} | \gamma \delta \rangle
            \hat{c}_\alpha^{\dag} \hat{c}_\beta^{\dag}
            \hat{c}_\delta \hat{c}_\gamma \; ,
\label{intraband}
\end{equation}
with single-particle energies $E^e_\alpha$ and
matrix elements of the Coulomb interaction between electrons
$ \langle \alpha \beta | V_{ee} | \gamma \delta \rangle$.
In the following, we will consider nanostructures exposed 
to strong magnetic fields, including the Zeeman spin splitting.
For present purposes, we restrict ourselves to harmonic-confining 
potentials.

\vspace{0.2in}
\noindent
{\large\bf AFMC METHOD}
\vspace{0.2in}

To calculate expectation values in AFMC, we make use of the 
Euclidian-time many-body propagator
$\hat{U}=\exp(-\beta\hat{H}_{ee})$, where $ \beta \equiv T^{-1}$
is interpreted as the inverse of the temperature.
For example, we calculate the expectation value of some observable 
$ \hat{\Omega}$
in a canonical ensemble, with fixed particle number $N$,
as
\begin{equation}
\langle \hat{\Omega} \rangle = 
 {{{\hbox{Tr}} [ \hat{P}_N \hat{U} \hat{\Omega}]}
\over{{\hbox{Tr}} [ \hat{P}_N \hat{U}] }}\;=
 {{{\hbox{Tr}}_N [\hat{U}\hat{\Omega}]}
\over{{\hbox{Tr}}_N [\hat{U}] }}\;,
\label{expectation}
\end{equation}
where the many-body trace is defined as $\hbox{Tr}\hat{X} \equiv
\sum_i \langle i | \hat{X}| i \rangle$ and the sum is over all
many-body states of the system.
If $\hat{N}$ is the number operator and $ \hat{P}_N 
= \delta (N-\hat{N})$ is the projector onto states with $N$ electrons, 
the canonical ensemble is defined by  $\hbox{Tr}_N \hat{X} 
\equiv \sum_i \langle i | \hat{P}_N \hat{X}| i \rangle$.

Beyond such static properties, this approach
allows one to obtain information about the dynamical response
of the system.
For operators $ \hat{\Omega}^\dagger $ and $ \hat{\Omega} $,
the response function $ R_\Omega (\tau) $ in the canonical
ensemble is defined as  
\begin{equation}
R_\Omega(\tau)
 \equiv 
{{\rm Tr}_N\, e^{-(\beta-\tau)\hat{H}_{ee}} \hat
\Omega^\dagger e^{-\tau\hat{H}_{ee}}\hat \Omega\over
{\rm Tr}_N\, e^{-\beta\hat{H}_{ee}}}  \\
 \equiv 
\langle\hat \Omega^\dagger(\tau)\hat \Omega(0)\rangle_N \; ,
\end{equation}
where  $ \hat{\Omega}^\dagger (\tau) 
\equiv  \exp(\tau \hat{H}_{ee}) \hat{\Omega}^\dagger 
\exp(-\tau \hat{H}_{ee} ) $
is the Euclidian-time Heisenberg operator.
Inserting complete sets of $N$-body eigenstates of $\hat{H}_{ee}$
($ \{ |i\rangle, | f | \rangle \})$ with energies $ E_{i,f}$ shows that
\begin{equation}
R_\Omega(\tau) ={1\over Z}\sum_{if} e^{-\beta E_i}
\vert\langle f\vert\hat \Omega\vert i\rangle\vert^2
e^{-\tau(E_f-E_i)} \; ,
\end{equation}
where $ Z = \sum_i \exp(-\beta E_i) $ is the partition function.
Thus, $R_\Omega (\tau)$ is the Laplace transform of the strength
function $S_\Omega (E)$:
\begin{eqnarray}
R_\Omega(\tau)& = & \int^\infty_{-\infty} e^{-\tau E} S_\Omega(E)dE\;; 
\\
S_\Omega(E)&=&  {1\over Z}\sum_{fi} e^{-\beta E_i}
\vert\langle f\vert \hat \Omega\vert i\rangle\vert^2 
\delta(E-E_f+E_i)\;.
\end{eqnarray}
It is important to note that we cannot usually obtain detailed
spectroscopic information (i.e., energies and wave functions)
from AFMC.  
Rather, we can calculate expectation values of operators in
the thermodynamic ensembles or the ground state \cite{kdl97}.

In the AFMC method, two-body interactions in $\hat{H}_{ee}$ are 
linearized through the Hubbard-Stratonovich (HS) 
transformation \cite{hstran}.
The difficult many-body evolution $U$ is replaced by a superposition 
of an infinity of tractable one-body evolutions, each in a different 
fluctuating external field, $\sigma$. 
Integration over the external fields thus reduces the many-body 
problem to quadrature, which is evaluated stochastically.

The many-body Hamiltonian can be written schematically as
\begin{equation}
\hat H=\varepsilon\hat {\cal O} +{1\over2}V\hat{\cal O}\hat{\cal O}\;,
\end{equation}
where $\hat{\cal O}$ is a density operator of the form $a^\dagger a$, $V$
is the strength of the
two-body interaction, and $\varepsilon$ a single-particle energy. 
In the full problem, there are many such quantities with various 
orbital indices that are summed over, but we omit them here for
clarity.

All of the difficulty arises from the two-body interaction.
If $\hat H$ were solely linear in
$\hat{\cal O}$, we would have a one-body quantum system, which is readily
dealt with. 
To linearize the evolution, we employ the Gaussian identity
\begin{equation}
e^{-\beta\hat H} =
\sqrt{\beta \mid V\mid \over 2\pi} \int^\infty_{-\infty}
d\sigma e^{-{1\over2}\beta \mid V\mid \sigma^2} e^{-\beta\hat h}\;\; ; \;\;
\hat h = \varepsilon \hat{\cal O} +s V\sigma\hat{\cal O}\;.
\label{gaussian}
\end{equation}
Here, $\hat h$ is a one-body operator associated with a $c$-number field
$\sigma$, and the many-body evolution is obtained by integrating the one-body
evolution $\hat U_\sigma\equiv e^{-\beta\hat h}$ over all $\sigma$ with a
Gaussian weight. The phase $s$ is $1$ if $V<0$ or $i$ if
$V>0$.
Equation (\ref{gaussian}) is easily verified by completing the square
in the exponent of the integrand and then doing the integral.

With an expression of the form (\ref{gaussian}), 
it is straightforward to write observables as the ratio of two integrals. 
For example, the canonical
expectation value (\ref{expectation}) becomes
\begin{equation}
\langle\hat \Omega\rangle_N=
{\int d\sigma e^{-{\beta\over2}\mid V\mid \sigma^2}{\rm Tr}_N\,
\hat U_\sigma\hat \Omega\over
\int d\sigma e^{-{\beta\over2}\mid V\mid\sigma^2}{\rm Tr}_N\, 
\hat U_\sigma}=
{\int d\sigma W_\sigma \Omega_\sigma\over
\int d\sigma W_\sigma}\; 
\label{ratio}
\;,
\end{equation}
where $W_\sigma = G_\sigma {\rm Tr}_N\, \hat U_\sigma$,
$G_\sigma = e^{-{\beta\over2}\mid V\mid \sigma^2}$, and
$ \Omega_\sigma =
({\rm Tr}_N\,\hat U_\sigma\hat \Omega) / ({\rm Tr}_N\,\hat U_\sigma)$.
Thus, the many-body observable is the weighted average (weight $W_\sigma$) of
the observable $\Omega_\sigma$ calculated in a canonical ensemble
involving only the one-body evolution $\hat U_\sigma$. 

An expression of the form (\ref{ratio}) has a number of 
attractive features. 
First, the problem has been reduced to quadrature. 
Second, all of the quantum mechanics 
is of the one-body variety, which scales simply with the square
of the number of single-particle states included in the calculation. 
The price to pay is treating the one-body problem 
for all possible $\sigma$.

Since $\hat{H}_{ee}$ contains many two-body terms that do not commute,
the evolution must be discretized, 
i.e., $\beta=N_t\Delta\beta$, before applying the HS transformation, i.e.,
\begin{equation}
Z_N = \hbox{Tr}_N e^{-\beta\hat{H}_{ee}}
 \rightarrow
 {\hbox{Tr}}_N
  \left[e^{-\Delta\beta\hat{H}_{ee}}\right]^{N_t} 
\rightarrow
\int{\cal D}[\sigma]G(\sigma){\hbox{Tr}}_N\prod_{n=1}^{N_t}
e^{\Delta\beta\hat{h}(\sigma_n)}\ ,
\end{equation}
where $\sigma_n$ are the auxiliary fields at a given imaginary
time-step $\Delta\beta$,
${\cal D}[\sigma]$ is the measure of the integrand, $G(\sigma)$ is a
Gaussian in $\sigma$, and $\hat{h}$ is a one-body Hamiltonian.
Dimensions of the integral can reach up to 10$^5$ for systems of
interest, and it is thus natural to use Metropolis random walk
methods to sample the space.
Because the numerical effort for AFMC scales only as a low power of
the problem size, very large (and hence more realistic) calculations
are possible.

The most significant challenge to applying the AFMC method
resides in overcoming the fermion-sign problem.
Algorithmic solutions to these problems are typically not robust
and are highly dependent on the fermion system to which they are
applied.
For example, systems whose interactions are purely attractive
are free of the sign problem \cite{kdl97}.

For our system of many electrons confined in a quantum dot, the
Coulomb forces are purely repulsive.
In computing expectation values of observables as in 
Eq.\ (\ref{expectation}), we will employ the finite-temperature 
methods discussed above.
At high temperatures, these methods will not suffer from a 
Monte Carlo sign problem, but at lower temperatures, we will 
encounter the sign problem.
Its most frequent manifestation occurs when the weight functions 
used for a Metropolis Monte Carlo evaluation of the integrals 
introduced by the Hubbard-Stratonovich 
transformation lose their positive-definite character \cite{kdl97}.
We are exploring methods to stabilize the AFMC against the sign
problem when the effective interactions include terms that are 
both attractive and repulsive \cite{Neuhauser,Zhang}.

Previously, 
we developed the AFMC method for the nuclear shell-model problem
and performed state-of-the-art calculations for a wide range of
nuclear properties, including response functions at finite 
temperature \cite{kdl97,dean98,radha97}.
The exciton density in the correlated nuclear system has
been described by an application of AFMC techniques to the nuclei
$^{40}$Ca and $^{42}$Ca in a very large model space \cite{dean99}. 

\vspace{0.2in}
\noindent
{\large\bf APPLICATION and RESULTS}
\vspace{0.2in}

We consider the problem of N electrons of effective mass $m^*$
in a plane, $(x,y)$, confined by an external parabolic potential,
$V(r) = \frac{1}{2} m^* \omega_O^2 r^2 $, and subject to a
strong magnetic field $\vec{B} = B_0 \vec{e}_z $. 
We consider the Zeeman splitting but neglect the 
spin-orbit interaction.
The Hamiltonian for such a system is 
\begin{equation}
\hat{H}_{ee} =
 \sum_i \left[
   \frac{\left( \vec{p}-\frac{e}{c}\vec{A} \right)^2}{2m^*}
 + V(r_i) 
 + \frac{g^* \mu_B}{\hbar} \vec{B} \cdot \vec{S}_i  \right] 
 + \sum_{i<j} \frac{e^2}{\varepsilon |\vec{r}_i - \vec{r}_j | }\;,
\end{equation}
where the vector potential is $\vec{A}(\vec{r}_i) = (B_0/2)(-y_i,x_i,0)$,
$g^*=0.54$, $\varepsilon=12.9$, $m^*=0.067 m_e$, and 
$\hbar \omega_0 = 3 meV$.


AFMC calculations require as basic inputs single-particle
energies, two-body matrix elements of the interaction,
and the initial many-body wavefunctions
which we obtain from Hartree-Fock (HF) calculations.
The HF method provides physically meaningful single-particle 
states including the exchange interactions exactly \cite{muller}.

\begin{figure}[t]
\centering
\epsfig{file=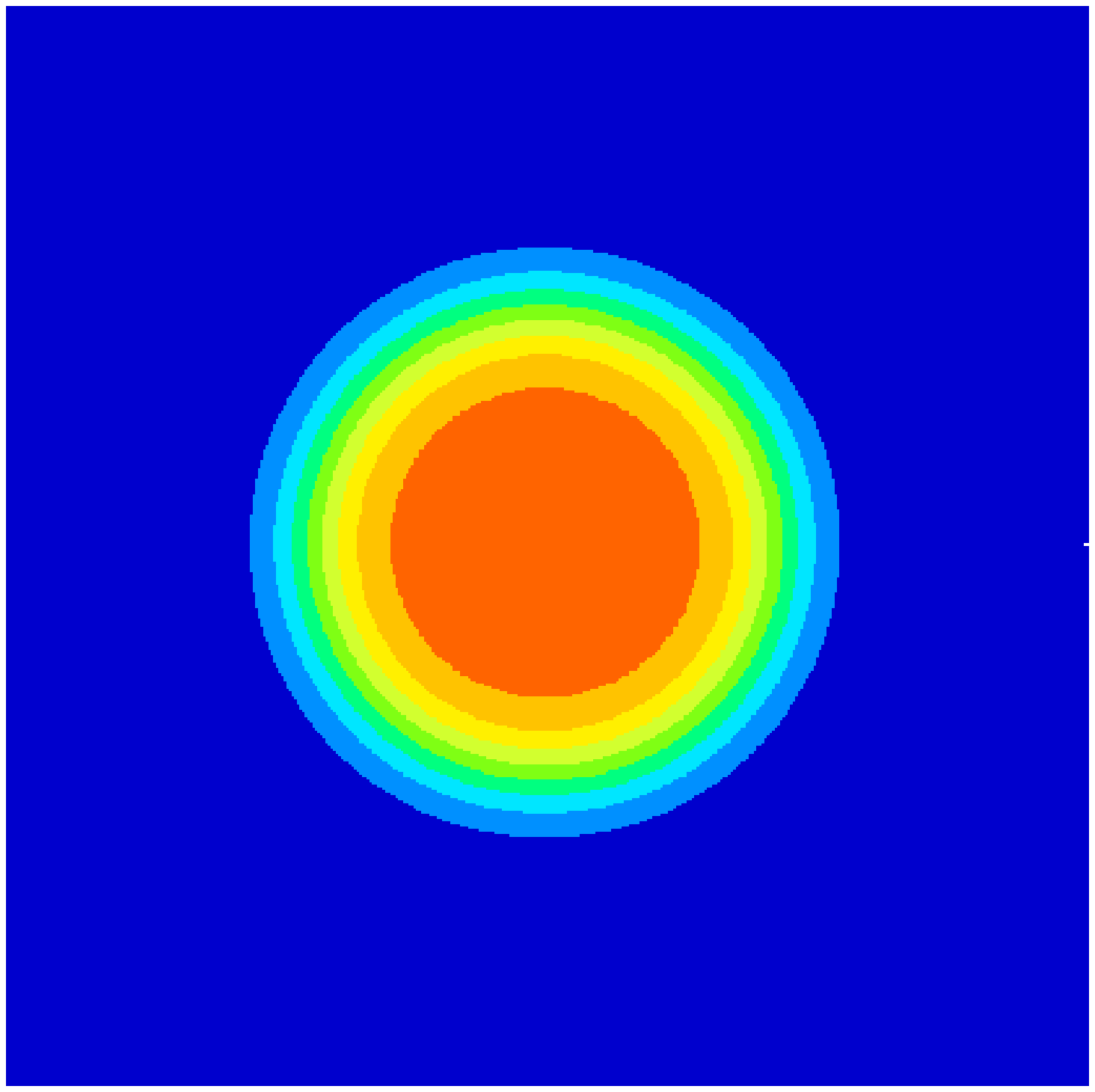,width=2.0in}
\epsfig{file=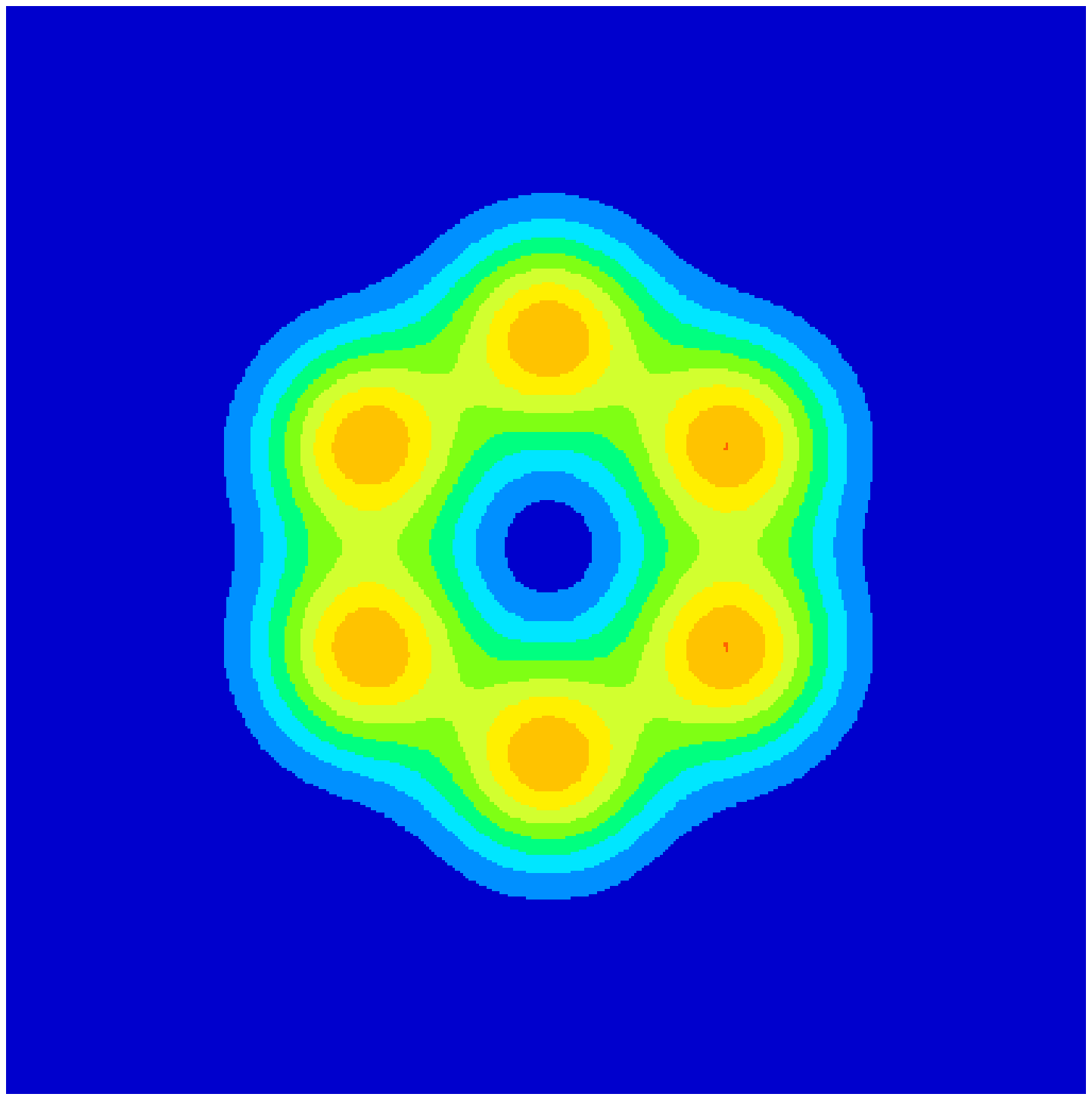,width=2.0in}
\epsfig{file=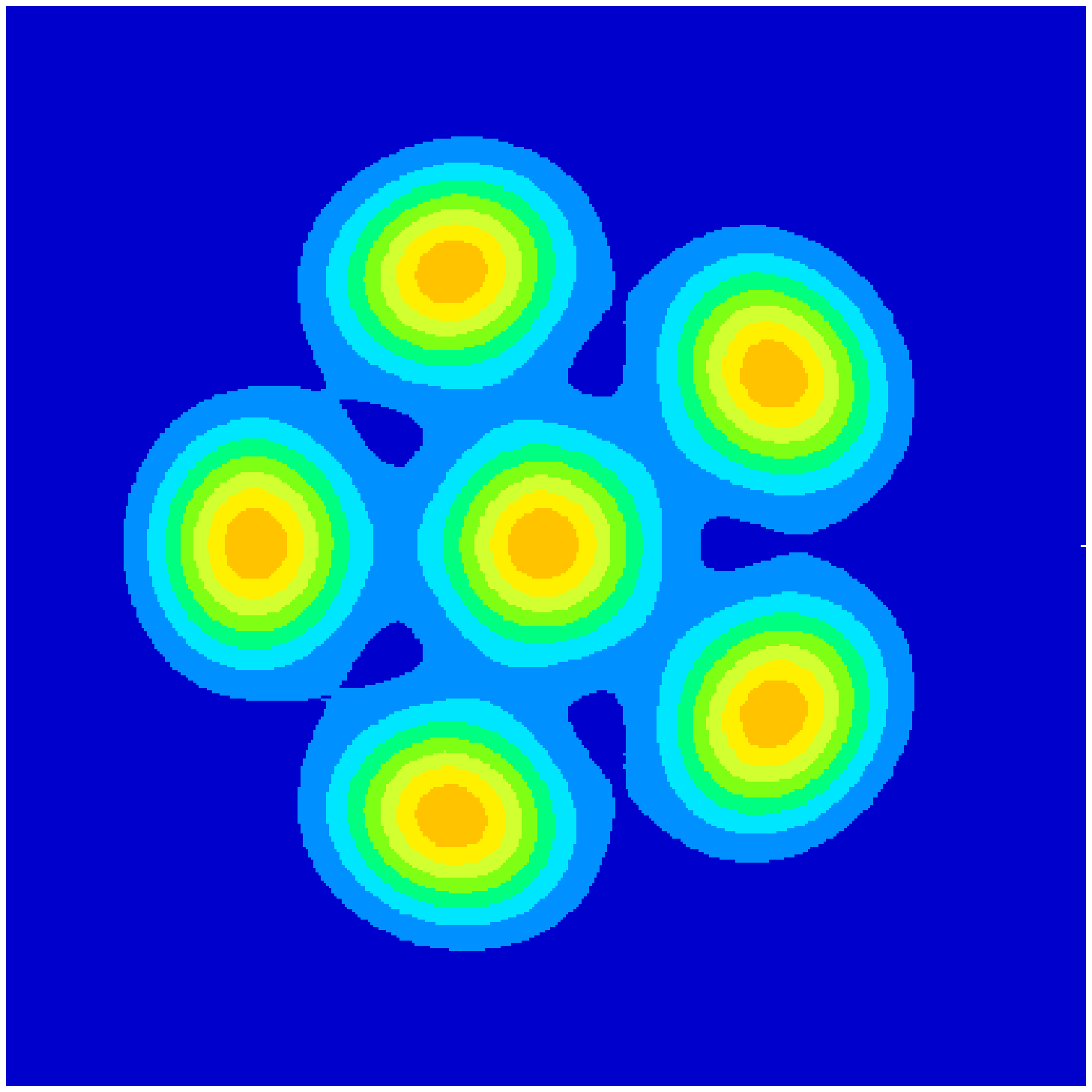,width=2.0in}
\caption{
(a) Maximum-density droplet;
(b) broken-symmetry phase, ``Wigner molecule''; and
(c) ``Wigner crystal''.
}
\vspace{-0.2in}
\label{density}
\end{figure}

Much experimental effort has focused on mapping the magnetic-field
dependence of the structure of semiconductor quantum dots 
by measuring the chemical potential via capacitance 
spectroscopy \cite{Ashoori}.
Cusps and steps in the chemical potential were found to clearly
separate different ranges of magnetic fields \cite{Ashoori,Oosterkamp}.
These features were identified with phase transitions in the charge 
density of the quantum dot. 
For increasing magnetic field, all electrons will become 
spin-polarized initiating the maximum density droplet (MDD)
phase \cite{Ashoori} [see Fig.\ (\ref{density}a)].
In this phase, the density is constant and homogeneous at the 
maximum value that can be reached in the lowest Landau level.
The stability of the MDD is determined by a competition between the
kinetic and external confinement and the Coulomb repulsion between
electrons.
For increasing magnetic field, the charge-density distribution
of the droplet reconstructs \cite{Chamon} with a ring of electrons
breaking off from the MDD phase (see Fig.\ (\ref{density}b)).
This edge reconstruction has been shown via mean-field calculations
to result from a rotational symmetry-breaking phase transition 
to a Wigner molecule or Wigner crystal phase \cite{muller,Reimann}
[see Figs.\ [\ref{density}b) and (\ref{density}c)].
These calculations are in good qualitative agreement with recent
experimental results \cite{Oosterkamp}.

\begin{figure}
\vspace{-1.2in}
\epsfig{file=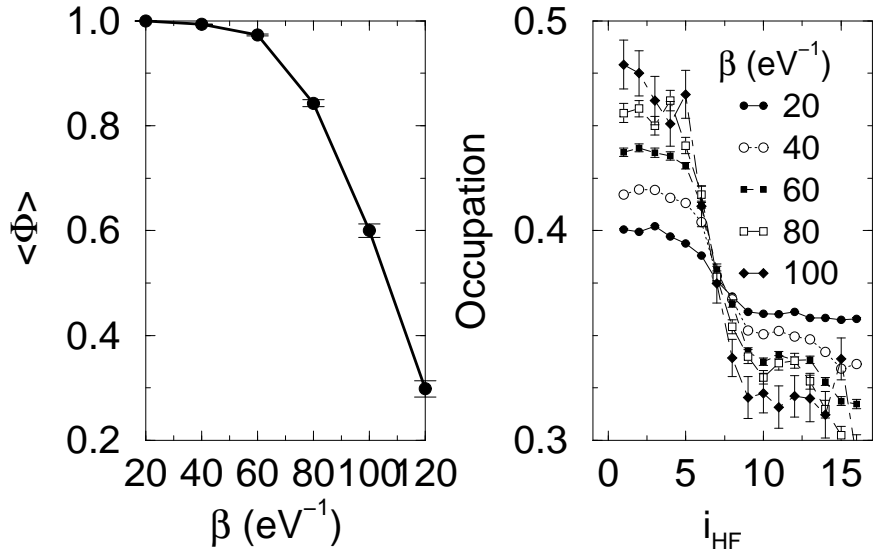,width=5.5in}
\caption{
Left: a) sign problem as a function of inverse temperature;
right: b) occupation of the HF states, $i_{HF}$ with decreasing temperature. 
}
\vspace{-0.2in}
\label{afmc}
\end{figure}

While we are in the initial stages of testing and 
calibrating our AFMC code, we do have several interesting
initial results. We begin with 
a canonical ensemble of $N=6$ electrons in a 
$B=4$~T magnetic field. The corresponding Hartree-Fock 
solution is shown in Fig.~(1a).
We first demonstrate in Figure (2a) the 
behavior of the Monte Carlo sign $\langle \Phi\rangle$ 
as a function of inverse temperature $\beta$. Note that
beyond approximately $\beta=80$~eV$^{-1}$, the sign begins to drop 
quickly. In order to obtain good evaluations of expectation
values beyond $\beta=100$~eV$^{-1}$, we need to calculate
many more Monte Carlo samples than here, where we 
have 2560 samples for each $\beta$. Figure (2b) shows the occupation of 
the various Hartree-Fock states at different temperatures. Thermal
excitation causes a degradation of the spin alignment, and thus
an occupation of both spin-up and spin-down levels. An interesting
near-term calculation will be to find the transition temperature
when the return of spin polarization occurs.

We are at the beginning of an exciting and challenging 
study of the optical properties of quantum dots. At the present,
we have a partially developed AFMC code with which to begin our
work. We shall include computations of E1 dipole-transition
operators in the code, and will investigate 
various techniques for overcoming the 
Monte Carlo sign problem for low-temperature studies. 

\vspace{0.2in}
\noindent
{\large\bf ACKNOWLEDGEMENTS}
\vspace{0.2in}

Research sponsored by the Laboratory Directed Research and Development
Program of Oak Ridge National Laboratory, managed by Lockheed Martin
Energy Research Corp. for the U. S. Department of Energy under Contract
No. DE-AC05-96OR22464.

\end{document}